# Digital cytometry: extraction of forward and side scattering signals from holotomography

JAEPIL JO,[1,2] HERVE HUGONNET,[1,2] MAHN JAE LEE,[2,3] AND YONGKEUN PARK[1,2,4,*]

[1]*Department of Physics, Korea Advanced Institute of Science and Technology, Daejeon 34141, Republic of Korea*
[2]*KAIST Institute for Health Science and Technology, KAIST, Daejeon, Republic of Korea*
[3]*Graduate School of Medical Science and Engineering, Korea Advanced Institute of Science and Technology (KAIST), Daejeon, 34141, South Korea*
[4]*Tomocube Inc., Daejeon, 34109, South Korea*
*[*]yk.park@kaist.ac.kr*

**Abstract:** Flow cytometry is a cornerstone technique in medical and biological research, providing crucial information about cell size and granularity through forward scatter (FSC) and side scatter (SSC) signals. Despite its widespread use, the precise relationship between these scatter signals and corresponding microscopic images remains underexplored. Here, we investigate this intrinsic relationship by utilizing scattering theory and holotomography, a three-dimensional quantitative phase imaging (QPI) technique. We demonstrate the extraction of FSC and SSC signals from individual, unlabeled cells by analyzing their three-dimensional refractive index distributions obtained through holotomography. Additionally, we introduce a method for digitally windowing SSC signals to facilitate effective segmentation and morphology-based cell type classification. Our approach bridges the gap between flow cytometry and microscopic imaging, offering a new perspective on analyzing cellular characteristics with high accuracy and without the need for labeling.

## 1. Introduction

Flow cytometry is a rapid and highly efficient analysis method that relies on the scattering signals of flowing cells, and it has long been a cornerstone in clinical applications and diagnostics [1-5]. Its major advantage lies in its ability to assess individual cells with much higher throughput compared to manual microscopic observations, making it invaluable for large-scale studies. Flow cytometry integrates microfluidics with scattering signal detection at two angles to analyze particles or cells (Fig. 1(a)). The first scattering parameter is forward scatter (FSC), which measures the intensity diffracted at small angles, depending on the size of the particle. The second is side scatter (SSC), which captures the intensity of the beam diffracted perpendicularly to the illumination, reflecting the structural complexity of the cell, called cell granularity [6]. In addition to scattering signal, multiple fluorescence signals can also be obtained to increase specificity.

To gain more detailed morphological information, imaging flow cytometry has been developed and widely adopted. This technique combines the high throughput of traditional flow cytometry with the advantages of single-cell image acquisition typically associated with microscopy (Fig. 1(b)) [7-10]. Imaging cytometry captures fluorescence or bright-field images of cells, and these signals can be further enhanced by combining them with artificial intelligence to effectively extract cell-type-specific information [11-13]. Bright-field or phase contrast microscopic techniques enable label-free imaging of cells, but with the limited imaging contrast (Fig. 1(c)) [14-16]. To address this, quantitative phase imaging (QPI) techniques have recently been integrated into flow cytometry setups [17-21]. QPI measures phase delay maps of cells, enhancing imaging contrast while maintaining cellular viability due to its label-free nature.

While microscopic imaging techniques provide direct spatial information about cells, the use of fluorescence probes allows for molecular-specific information about subcellular organelles or proteins [22]. However, microscopic images alone do not offer insights into scattering signals. One can only qualitatively infer cell size and complexity from micrographs, whereas FSC and SSC signals collected in flow cytometry do not provide a complete description of cellular morphology. Despite the long history of these two distinct techniques—microscopic cellular imaging and cytometry scattering signals—a comprehensive investigation combining scattering and imaging of cells has not yet been fully explored.

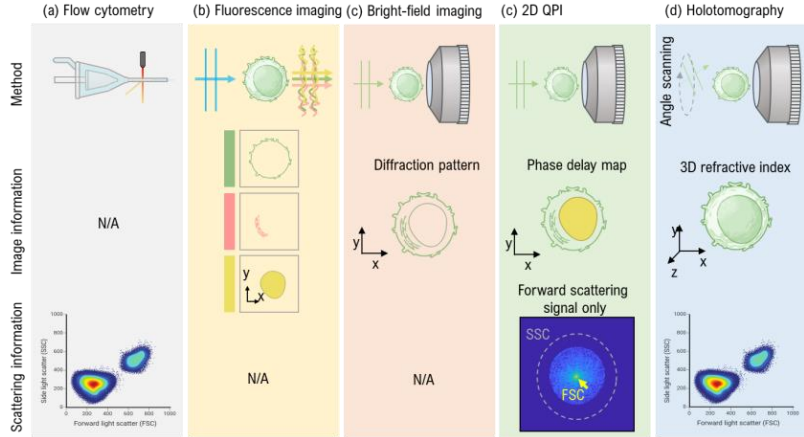

Fig. 1. Several methods to obtain the scattering singles and image data. (a) Flow cytometry measures the scattering signals, but the images are not recorded. (b) Fluorescence image-based cell analysis utilizing the labeled image. It needs the staining and cannot retrieve the scattering signals. (c) 2D QPI measures the 2D holographic image of the sample, and the scattering signals can be retrieved by FTLS method. Because the maximum angle that this method gathers is limited to NA of objective lens, it cannot retrieve the side scatter of the sample. (d) Holotomography restore the 3D refractive index of the sample. The scattering signals over the angle of 90° are also obtained with proposed method.

In this work, we aim to bridge these two separately developed techniques by quantitatively extracting FSC and SSC signals from 3D refractive index (RI) tomograms of cells (Fig. 1(e)). We present a computational method for extracting both FSC and SSC signals from individual cells using 3D RI tomograms experimentally measured through holotomography (HT). Holotomography is a 3D QPI technique that enables the reconstruction of 3D RI tomograms from multiple 2D light field maps obtained with various illumination angles [23]. By using Fourier transform light scattering (FTLS) [24-26], we retrieve 2D light scattering spectra from these measured 2D field images. The light scattering signals corresponding to either forward or side scattering angles are digitally reorganized and presented at the individual cell level. Due to the high numerical apertures of both the condenser and objective lenses, holotomography allows for the collection of scattering signals beyond the traditional 90-degree angles. We demonstrate the extraction of FSC and SSC from various types of samples, including microspheres and white blood cells. Additionally, we present the extraction of scattering signals corresponding to specific angles as required by HT.

## 2. Methods

### 2.1 Fourier transform light scattering

The Fourier transform light scattering (FTLS) is a technique to obtain the angle-resolved light scattering (ARLS) from an optical field measured at the image plane [24], using QPI techniques. The far-field scattering intensity can be calculated using a 2D spatial Fourier transformation,

$$\left|E(k_x, k_y)\right|^2 = \left|\frac{1}{2\pi} \iint E(x,y)e^{-i(k_x x + k_y y)} dx dy\right|^2, \qquad (1)$$

where $k_x$ and $k_y$ are spatial frequencies: $k_x = 2\pi n_m(\sin\theta_x/\lambda)$, $k_y = 2\pi n_m(\sin\theta_y/\lambda)$ (Ref. [27]), $\theta_x$ and $\theta_y$ are scattering angles for the $x$ and $y$ axis, and $n_m$ is the RI of a medium surrounding a sample.

Compared to the conventional mechanical scanning methods, such as a goniometer[28], FTLS has several advantages: (i) extraction of scattering signals from individual imaging objects and (ii) high signal-to-noise ratio. Recently FTLS has been utilized for the study of colloidal particles, blood cells, and bacteria [25, 29-32]. Although FTLS with 2D QPI provides access to both the image and ARLS, the access to SSC using FTLS remains challenging due to the limited NA of an objective lens (Fig. 1(c)).

*2.2 Experimental setup*

The experimental amplitude and phase images were recorded with a HT setup (HT-2H, Tomocube Inc., South Korea), and a schematic of the measurement setup is depicted in Fig. 2(a). This HT setup utilizes a Mach–Zehnder interferometer that captures off-axis holograms using a diode-pumped solid-state laser with a wavelength of 532 nm [33]. The coherent light source is split into two beams: one serves as an off-axis reference, and the other is used to illuminate the sample. The illumination angle of the beam impinging onto a sample is modulated by using a digital micromirror device (DMD, DLPLCR6500EVM, Texas Instruments Inc., USA) [34, 35]. The incident beam is directed onto the sample plane through a 4-$f$ telecentric imaging system composed of a tube lens and a water immersion condenser lens (numerical aperture (NA) = 1.2, UPLSAPO 60XW, Olympus Inc., Japan). An objective lens (NA = 1.2, water immersion, UPLSAPO 60XW, Olympus Inc., Japan) and a tube lens ($f$ = 175 mm) collect the transmitted beam and project the transmitted image to the image plane. Both the sample and the reference beam interfere at an image plane, and generate spatially modulated interference patterns, which are captured using an image sensor (BFS-U3-28S5M-C, FLIR, USA). The amplitude and phase image are retrieved from the measured interferogram using a field retrieval algorithm [36].

*2.3 Sample preparation*

We drew peripheral blood from healthy donors into tubes coated with ethylene-diamine-tetra acetic acid (EDTA). We then separated peripheral blood mononuclear cells (PBMCs) using density gradient centrifugation. Following this, we extracted monocytes with the pan monocyte isolation kit (Miltenyi Biotec, Germany), as per the provided instructions. Specifically, we diluted 10mL of whole blood equally with the autoMACS® Rinsing Solution (Miltenyi Biotec, Germany), which includes bovine serum albumin. We layered 20 mL of this diluted mixture over 15 mL of Histopaque®-1077 (Sigma-Aldrich, USA) and centrifuged it at 700 g for 30 minutes at room temperature. We then gently gathered the mononuclear layer and washed it two times. The gathered PBMCs were mixed with MACS buffer and specific antibodies bound with magnetic beads (Miltenyi Biotec, Germany). Magnetic cell separation was performed to isolate non-labeled cells, thereby enriching the sample of monocytes. We used the same methods to extract T cells using the Pan T cell isolation kit (Miltenyi Biotec, Germany). We then dispersed the gathered monocytes in a complete medium made of RPMI-1640 (Gibco®, USA), enriched with 10% Fetal bovine serum (Gibco®, USA) and 1% penicillin/streptomycin (Sigma-Aldrich, USA).

To extract granulocytes, we layered density gradient media (Histopaque 1119 and Histopaque 1077, ThermoFisher, USA), on top of which we carefully added a diluted mix of whole blood and phosphate-buffered saline (PBS, ThermoFisher, USA) with 2% Fetal Bovine Serum (FBS, ThermoFisher, USA). After centrifuging at 800 g for 30 minutes without a break, we found granulocytes in the middle layer of the gradient media. These granulocytes were rinsed twice with 10 mL of PBS and centrifuged at 200 g for 10 minutes. Finally, we resuspended the collected cell concentrate in 5 mL of PBS.

The Internal Review Board (IRB) at KAIST approved and reviewed this study (Approval No. KH2017-004). Our methods complied with the Helsinki Declaration from 2000, and we secured informed consent from all subjects.

## 3. Results and discussion

### 3.1 Extraction of FSC and SSC using holotomography

In wave physics, both scattering and imaging originate from the same underlying information— namely, the 3D distribution of scattering potentials or RI distributions. The 3D RI distribution governs light-matter interactions, and both scattering and imaging represent different basis transformations of this information, whether in spatial or frequency coordinates. Essentially, they span the same information but in different dimensions.

To extract both FSC and SSC from HT, we employed an optical setup based on Mach-Zehnder interferometry, which is equipped with a DMD (Fig. 2(a), see Methods). The coherent laser source, with a central wavelength of 532 nm, generates a spatially filtered plane wave that is split into two arms. One arm serves as the sample beam, which is spatially modulated using the DMD before impinging on the sample, while the other arm acts as a reference beam that reaches the image sensor. The light diffracted from the sample is collected using a high NA lens and projected onto the image plane. At this image plane, the sample and reference beams interfere with a slight tilt angle, forming a spatially modulated hologram pattern. The DMD systematically controls the illumination angle of the beam projected onto the sample.

Under normal illumination, the light field image of a sample is retrieved from a measured hologram (Fig. 2(b)). From the measured hologram, the both the amplitude and phase image of the sample are retrieved using a field retrieval algorithm. The maximum angle of ARLS is limited by the acceptance angle $\theta_{max}=sin^{-1}(NA/n_m)$ of the objective lens depicted as green fanwise areas in Fig. 2(b). For our setup $\theta_{max}$=63 degree, meaning the objective lenses cannot collect the side scattering signal generally representing light scattered at 90° from the incident beam. To increase this maximum angle, we employed oblique angle illuminations. When the same lenses are utilized for both the condenser and the objective lens, the illumination angle can coincide with the maximum acceptance angle of the objective lens. As a result, the theoretical maximum angle of ARLS can potentially be extended to $2\theta_{max}$=126 degree.

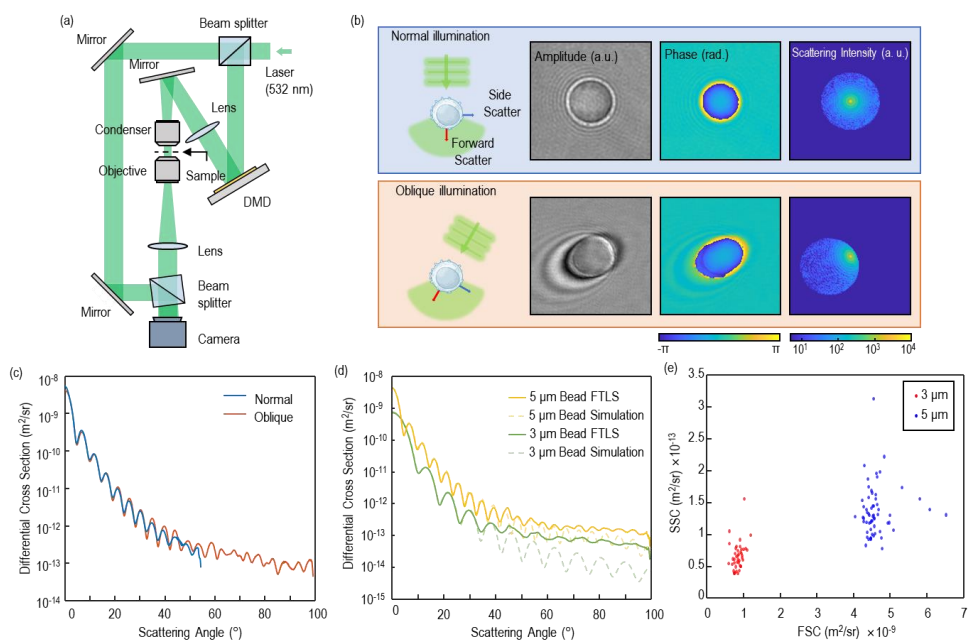

Fig. 2. (a) Schematic representation of the HT setup. (b) Schematics of forward, side scatter signals and collective angle of objective lenses in normal and oblique illuminations. Measured amplitudes and phase of microbeads under normal and oblique illuminations. Fourier transform images of electric fields under normal and oblique illumination. The log scale color bar is adopted for Fourier transform images. (c) Obtained angle-resolved light scattering represented as a differential cross section. The blue line and orange line show the ARLS under normal illumination and oblique illumination, respectively. (d) Averaged ARLS in various illumination angles. Yellow and green lines are for silica beads with diameters of 5 μm and 3 μm. Solid lines represent the experimental ARLS using FTLS, and dashed lines are the result of the Mie scattering simulation. (e) The scatter plot illustrates the FSC and SSC for two types of beads. The red points represent the scatter signals from 3 μm beads, while the blue points indicate those from 5 μm beads. 42 of 3 μm beads and 61 of 5 μm beads are used.

A 5-μm-diameter silica bead is used to demonstrate our idea. The amplitudes and phases of the beam transmitted through beads and the Fourier transforms of light fields are presented in Fig. 2(b). The NA circle of the Fourier-transformed image is displaced to the side under oblique illumination. This translation is responsible for the extension of the maximum angle of ARLS.

The results of FTLS for normal and oblique illumination are shown in Fig. 2(c). The ARLS is represented by the quantitative value known as the differential cross section, unlike previous studies that expressed it as an arbitrary unit. The differential cross section is calculated by multiplying a factor of $(n_m/\lambda)^2 \Delta x^4$ to the mean intensity of each pixel in concentric circles where $\Delta x$ is the pixel size [37]. The differential cross section for normal illumination, represented by the blue line, has a limited range of up to 54° due to the numerical aperture of the objective lens. Although the theoretical maximum angle of our objective lens (NA = 1.2, water immersion) is 64°, the practical angle of acceptance was found to be smaller. For an oblique illumination at 45°, the angle range of ARLS is extended to over 90°. Despite the effects caused by the oblique illumination, such as sample rotation and asymmetrical signal collection, ARLS signals under both normal and oblique illuminations produce identical results which is expected due to the spherical symmetry of microbeads (Fig. 2(c)) [38].

*2.2 Comparison with Mie scattering theory*

We compared experimental data to theoretical predictions for microbeads. Figure 2(d) illustrates both the expected ARLS based on Mie theory and the measured ARLS for microbeads with diameters of 5 μm and 3 μm. The ARLS was determined by averaging signals from one normal illumination and 25 azimuthally scanned oblique illuminations. The parameters for the theoretical values were optimized by fitting the Mie scattering curve to the experimental ARLS data, aiming to achieve the maximum correlation coefficient within the angular range of 0° to 35°. For the 5 μm bead, the determined parameters were a diameter of 5.12 μm and a refractive index of 1.433. In comparison, the manufacturer's specified diameter is 5.00 μm, with a standard deviation (STD) of 0.18 and a coefficient of variation of 3.6%. The 3 μm bead was fitted with a diameter of 2.91 μm and a refractive index of 1.429 and the manufacturer's specification is 3.01 ± 0.12 (STD) μm with a coefficient variation of 4.1% in diameter. The diameters of the two microbeads, as estimated through coefficient fitting, align with the manufacturer's specifications. Additionally, the estimated refractive indices are consistent with the typical values of 1.42 to 1.43 for silica microbeads [39, 40].

In figure 2(d), the experimental and theoretical FTLS for 5 and 3 μm diameter silica beads are represented by solid and dashed lines, respectively. Both theoretical and experimental signal precisely match up to 40°. However, beyond this angle, the experimental signal shows higher values than theoretically predicted. This discrepancy is primarily due to the noise, which is more dominant at higher angles due to weaker sample signals.

*2.3 Forward and side scatter of microbeads*

We extracted the forward and side scatter signals of numerous microbeads from ARLS data as the differential cross section at 0° and 90°. On the FSC-SSC plot, the scattering signals for beads with diameters of 3 μm and 5 μm are marked with red and blue dots, as depicted in Fig. 2(e). The two types of beads are distinctly separated, which aligns with our expectation that bead types can be distinguished in scatter plot using FTLS. The forward scatter signals are more concentrated when compared to the side scatter because of the better signal-to-noise at the low angle. This result is consistent with conventional flow cytometry [41, 42].

In conventional flow cytometry, FSC and SSC signals are captured using different detectors and voltage settings [43]. A photodiode is typically used for FSC detection, whereas a photomultiplier tube is chosen for SSC. This approach is due to the need for high dynamic range in flow cytometry because of the very small scattering cross-section characteristic of single cells. The detection structures vary across different equipment, and the voltage settings depend on the type of samples being analyzed. These variations hinder the ability to unify results across different setups and measurement conditions. In our method, since the scattering intensity is quantified using the differential cross section based on the scattering potential of the samples at a given wavelength, it enables us to compare data measured in different optical setups. This is one of the strengths of the FSC-SSC plot when compared to FTLS.

*2.4 Scattering signals of human white blood cells*

We now proceed to apply our method to white blood cells (WBCs) because WBCs are one of the representative flow cytometry samples widely used for diagnosis. Utilizing three distinct types of WBCs, we demonstrate the capability of our methodology to simultaneously reconstruct the RI tomogram (Fig. 3(a)) and extract the quantitative scatter plot (Fig. 3(b)). Using an optical diffraction tomography algorithm [44, 45], we reconstructed the 3D RI tomograms from the same 2D holographic images used in FTLS. In Fig. 3(a), typical RI tomograms for three samples of each of the WBC types are displayed. Among these, the lymphocytes are smaller than the other two cell types leading to a weaker FSC signal, while the neutrophils are distinguished by their numerous granules leading to a stronger SSC signal. These unique morphological traits lead to differences in the magnitude of ARLS in Fig. 3(b). Although not directly comparable to conventional flow cytometry, results show similar trends [46]. Each type of human white blood cell is roughly gathered, and the trends in the strength of scattering signals are similar to the scatter plot of flow cytometry; Lymphocytes have the smallest FSC and SSC, and granulocytes have the largest side scatter signal. In the plot, each type of human white blood cell clusters together, showing a scattering signal strength trend that aligns with traditional flow cytometry scatter plots. Specifically, lymphocytes exhibit the smallest FSC and SSC, while granulocytes have the largest side scatter signal. 115 lymphocytes, 146 monocytes, and 42 neutrophils are included in Fig. 3(b).

The FSC-SSC plot is depicted in Fig. 3(b). FSC and SSC values are calculated by averaging the ARLS angles from 3° to 9° and from 87° to 93°, respectively. This approach is adopted because the region very close to 0° is typically obstructed for forward scattering to obtain a value more proportional to sample size. Most biological cells do not show spherical symmetry. In our method, averaging the scattering signal for different illumination and detection angles could give a statistically more significant value than conventional flow cytometry.

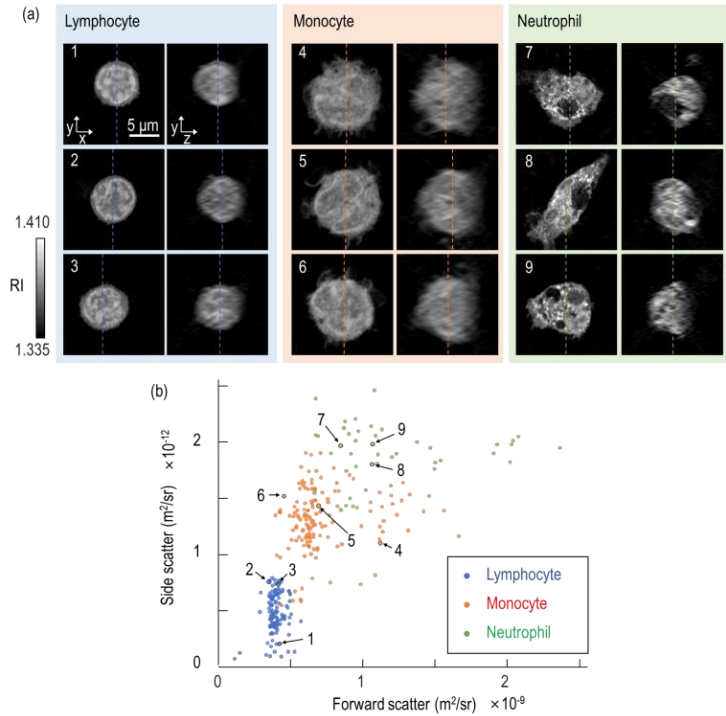

Fig. 3. (a) 3D refractive index distribution of human blood cells, with three samples illustrated for each type of WBC. The lateral resolution is 110 nm. (b) Forward and side scatter plot of WBCs, including 115 lymphocytes, 146 monocytes, and 42 neutrophils. Marked points correspond to the scattering signals of cells, numbered as shown in Fig. 3(a).

## *2.7 Scattering signal extraction in arbitrary angles*

The FTLS-based analysis enables the creation of scatter plots at arbitrary angles. To differentiate cell types, we can select the most appropriate angles other than the conventional forward and side scatter. We repeated the data analysis from section 2.4 with intermediate scatter (ISC) instead of the forward scatter signal. We experimented with three different ISC angle ranges of 27–33°, 42–48°, and 57–63°. Monocytes and neutrophils are much more separated in the ISC-SSC plot (Fig. 4) whereas the two types of WBC are partially overlapped in the FSC-SSC plot (Fig. 3(b); 57–63°). This feature is attributed to the fact that the intensity of scattered light reflects the morphology of the cells at different angles. Small angle scattering

(0.5°–5°) generally depends on the cell size and RI, and large angle scattering (15°–150°) is influenced by the internal granularity, organelles and surface roughness of a cell[47]. The enhanced separation of these two types in ISC-SSC plots is consistent with our knowledge. The significant distinction between monocytes and neutrophils primarily arises from the differences in the number of intracellular granules rather than cell size.

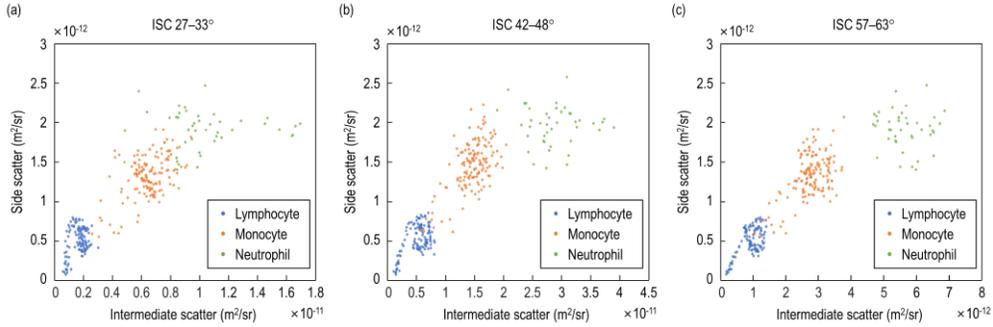

Fig. 4. The ISC-SSC plots for three types of white blood cells. 27–33°, 42–48°, and 57–63° are used as ISC angles in (a)–(c), respectively.

## 4. Conclusion

Flow cytometry remains a critical tool in diagnostics, cell cycle analysis, and immunophenotyping due to its ability to distinguish particles based on forward and side light scattering, which correlates with particle size and structural complexity. This paper has explored the relationship between scattering signals measured in flow cytometry and HT. By utilizing off-axis holography, quantitative phase imaging was achieved, allowing for the retrieval of both phase and amplitude data. To overcome the angle limitations imposed by the NA of the objective lens, we employed oblique illuminations within a HT setup.

The reliability of the present method was demonstrated by comparing experimentally recorded ARLS with theoretical calculations based on Mie theory for microbeads. Furthermore, the study successfully extracted forward and side scattering signals from ARLS at 0° and 90°, enabling the clear differentiation of bead types in the FSC-SSC plot.

Our methodology for deriving FSC and SSC from ARLS stands out by allowing the inclusion of signals from arbitrary angles as ISC. In our research, we have introduced this concept of ISC alongside the traditional SSC-FSC plot for the analysis of human white blood cells, thereby expanding the analytical capabilities in cell characterization. The FSC-SSC plot yielded results consistent with our expectations, and the ISC-SSC plots containing angle ranges of 27–33°, 42–48°, and 57–63° were able to separate the types of cells more clearly.

Although each method—flow cytometry and imaging—has traditionally been limited to analyzing either scattering signals or image data, our study could serve as a bridge between these two modalities. Recently the advent of imaging flow cytometry has merged the rapid analysis capabilities of flow cytometry with detailed single-cell imaging traditionally reserved for microscopy [48]. However, it is important to acknowledge that the current method's limitations, such as the lower SNR at higher scattering angles and reduced throughput, must be addressed for it to be widely applicable in clinical cytometry. Improving SNR by employing more sensitive image sensors with larger bit-depth capabilities, and utilizing zero-padding in the FTLS process to reduce noise (as per Parseval's theorem), could help overcome these challenges [49].

The scalability of this technique in clinical environments will also depend on improving throughput. By integrating this oblique-angle single-shot holography approach with imaging flow cytometry and automating data processing, high-throughput measurements could become feasible. This would allow the method to compete with conventional flow cytometry in terms of speed and applicability in large-scale diagnostics.

Our method shows potential to extend the maximum angle of FTLS beyond the NA of the objective lens, even with the use of just one oblique illumination. Integrating this oblique-angle single-shot holography approach with imaging flow cytometry could result in significant improvements in throughput and understanding of scattering phenomena.

While our results are promising, it is crucial to consider the broader implications of this work in the context of conventional flow cytometry, particularly regarding SNR, throughput, and scalability for large-scale clinical applications. In conventional flow cytometry, SNR is optimized using specialized detectors, whereas our method, which relies on FTLS and holotomography, faces challenges at higher scattering angles due to increased noise. Addressing these SNR limitations is essential for achieving comparable performance to conventional techniques. Additionally, the current lower throughput of our method, due to the complexity of data acquisition and processing, may limit its applicability in high-throughput clinical environments. Enhancing throughput through automation and faster computational algorithms will be vital for scaling this method to large-scale applications.

The method can be further extended by combining it with fluorescence signals [50, 51], allowing for correlative measurements of 3D RI tomograms and corresponding fluorescence data. This integration could enable a wide range of flow cytometry analyses, from basic single-image characteristics to advanced machine learning and deep learning methods, further expanding the capabilities and applications of this approach. The capabilities of imaging flow cytometry range from the utilization of basic single-image characteristics to advanced machine and deep-learning methods [7, 52, 53].

In summary, the present method offers advancements in combining scattering and imaging data and has the potential to complement or even enhance conventional flow cytometry in various research and clinical settings.

**Funding.** This work was supported by National Research Foundation of Korea (2015R1A3A2066550, 2022M3H4A1A02074314), Institute of Information & communications Technology Planning & Evaluation (IITP; 2021-0-00745) grant funded by the Korea government (MSIT), KAIST Institute of Technology Value Creation, Industry Liaison Center (G-CORE Project) grant funded by MSIT (N11220014, N11230131), Tomocube Inc.

**Disclosures.** H. Hugonnet, M. J. Lee, and Y. K. Park have financial interests in Tomocube Inc., a company that commercializes optical diffraction tomography and quantitative phase imaging instruments and is one of the sponsors of the work.

**Data availability.** Data underlying the results presented in this paper are not publicly available at this time but may be obtained from the authors upon reasonable request.